\documentclass[aps,prl,twocolumn,superscriptaddress,showpacs]{revtex4}
\usepackage{amsmath}
\usepackage{graphics}
\usepackage{graphicx}
\usepackage{epsfig}
\usepackage{amsmath}
\usepackage{amsfonts}
\usepackage{amssymb}

\begin{document}

\title{Spin wave instabilities and field induced transitions in heavy
fermions}
\author{M. A. \surname{Continentino}}
\email{mucio@if.uff.br}
\affiliation{Instituto de F\'{\i}sica, Universidade Federal Fluminense, Campus da Praia
Vermelha, Niter\'{o}i, 24210-340, RJ, Brazil}
\date{\today}

\begin{abstract}
We study phase transitions in heavy fermion systems due to spin-wave
instabilities. One motivation is to determine the changes in the
spin-wave parameters of a magnetically ordered heavy fermion system
as it approaches a quantum critical point (QCP) by applying
pressure.  The other more actual is to provide an alternative
approach, based on spin-wave instabilities, for the magnetic field
induced transitions recently observed in antiferromagnetic heavy
fermion materials.
\end{abstract}


\maketitle

\section{Introduction}

Pressure and magnetic field have been used as external parameters to drive
heavy fermion materials through different quantum critical points \cite%
{review}. The theory of the heavy fermion quantum criticality has been
developed starting on the paramagnetic side of their phase diagram \cite%
{mucio}. In the ordered magnetic phase, usually with
antiferromagnetic (AF) long range order, most studies have
concentrated in determining the shape of the critical line as a
function of pressure ($P$) and magnetic field ($H$), in particular
when $T_{N}\rightarrow 0$ \cite{review}. This gives information on
the important \emph{shift} exponent which is determined by the
critical exponents associated with the QCP. However, in the broken
symmetry phase, the magnetic excitations are spin-waves and a
relevant question concerns what happens to these modes as the
quantum critical point (QCP) is approached from the ordered side
\cite{suzana}. More specifically, we ask what happens to the
spin-wave parameters, its gap and stiffness as, for example,
$T_{N}(P)$ goes to zero, i.e., $P$ approaches $P_C$ the critical
pressure? The most direct way to answer this question would be to
perform neutron scattering experiments, but these are non trivial.
There are other less direct strategies to follow these modes as the
system approaches a QCP. One of these is the electrical resistivity.
Since magnons scatter the conduction electrons of the correlated
metal, the temperature dependent resistivity has a component which
is due to electron-magnon scattering. The temperature dependence of
this component yields direct information on the spin-wave gap and
stiffness \cite{suzana}.

\begin{figure}[th]
\centering \includegraphics[angle=0,scale=0.90]{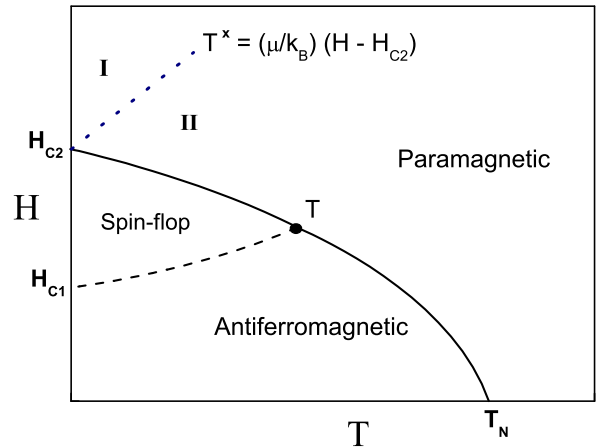}
\caption{(Color online) Phase diagram of an anisotropic
antiferromagnet in an external magnetic field in the easy axis
direction\protect\cite{keffer}. The three phases, antiferromagnet,
spin-flop and paramagnetic and the tricritical point (T) are
shown. The full lines are lines of second order phase transitions.
The dashed line is a line of first order transitions and the
dotted line is a crossover line dividing the paramagnetic region
in two regimes, as explained in the text. } \label{fig1}
\end{figure}

Recently, there have been many experimental studies on field induced
transitions in magnetic heavy fermion systems \cite{fieldinduced}. The
analysis of these experiments takes into account correlated electrons with
masses strongly renormalized by the presence of a QCP at the critical field $%
H_{C2}$ (see Fig.\ref{fig1}). However, field induced magnetic
transitions are associated with soft spin-wave modes. Consequently,
close to $H_{C2}$ these modes give an important contribution to the
thermodynamic properties and electrical resistivity, by scattering
the quasi-particles. These contributions should be included in an
analysis of this phase transition, besides that of other
quasi-particles.

The magnetic field and temperature dependent phase diagram of an
anisotropic antiferromagnet, with the external field $H$ applied
in the easy axis direction \cite{keffer}, is shown schematically
in Fig.\ref{fig1}. General features of this phase diagram are the
presence of a tricritical point and lines of first and second
order transitions. Heavy fermions are in general strongly
anisotropic and a relevant situation for these materials is the
\emph{metamagnetic case} where there is no intermediate spin-flop
phase at all and the system goes directly from the
antiferromagnetic phase to the
paramagnetic phase through a line of second order phase transitions, $%
H_{C2}(T)$.

\section{Resistivity of the anisotropic antiferromagnet in zero magnetic
field}

We start considering the case relevant for the study of heavy
fermions systems, \emph{driven by pressure} to a QCP. The spectrum
of hydrodynamic spin-wave excitations in the antiferromagnetic
phase, for $H << H_{C1}$ ($T<<T_N$) (see Fig.\ref{fig1}), is given
by,
\begin{equation*}
\omega (k)=-\mu H\pm \sqrt{\Delta ^{2}+Dk^{2}}
\end{equation*}%
where the gap $\Delta $ is proportional to the anisotropy (either exchange
or single ion). In the isotropic case and in the absence of a magnetic
field, the hydrodynamic magnons have a linear dispersion relation, $\omega
(k)=\sqrt{D}k$, differently from those in a ferromagnet, for which, $\omega
(k)=Dk^{2}$.

We consider first a metallic antiferromagnet in a simple cubic lattice with
no applied external field. Following Yamada and Takada \cite{yamada} we
write the electrical resistivity due to electron-magnon scattering as,
\begin{equation}
\rho (T)=CI_{AF}(H=0,T)
\end{equation}%
where $C$ is a constant. The quantity $I_{AF}(0,T)$ is given by
the following integral \cite{yamada},
\begin{equation}
I_{AF}(0,T)=\frac{\beta }{2}\int_{0}^{\infty }dk\frac{k^{2}}{\sinh ^{2}\frac{%
\beta \epsilon _{k}}{2}}
\end{equation}%
where $\omega(k)=\sqrt{\Delta ^{2}+Dk^{2}}$ is the dispersion relation of
the antiferromagnetic magnons. In the limit $\beta ^{-1}=k_{B}T<<\Delta $,
we deal with this integral changing variables and integrating by parts to
obtain,
\begin{equation*}
I_{AF}(0,T)=\frac{2(k_{B}T)^{2}}{D^{3/2}}y_{0}^{2}e^{-y_{0}}\int_{0}^{\infty
}e^{y_{0}(1-\cosh x)}\cosh 2xdx
\end{equation*}%
where $y_{0}=\Delta /k_{B}T$. The function $F(x)=e^{y_{0}(1-\cosh x)}$ is
such that, $F(0)=1$ and $F(x_{f})=e^{-1}$.The last condition implies $%
y_{0}(1-\cosh x_{f})=-1$, or $\cosh x_{f}=1+y_{0}^{-1}$. This function cuts
off the integral at $x_f$ and we can rewrite the equation above as,
\begin{equation*}
I_{AF}(0,T)=\frac{2(k_{B}T)^{2}}{D^{3/2}}y_{0}^{2}e^{-y_{0}}\int_{0}^{x_{f}}%
\cosh 2xdx
\end{equation*}%
which yields,
\begin{equation}
I_{AF}(0,T)=\frac{\Delta ^{2}}{D^{3/2}}e^{-\frac{\Delta }{k_{B}T}}\sinh
2x_{f}  \label{antiferrogap}
\end{equation}%
Since $\cosh x_{f}=1+y_{0}^{-1}=1+k_{B}T/\Delta $, $\sinh 2x_{f}=2\left(
1+y_{0}^{-1}\right) \sqrt{\left( 1+y_{0}^{-1}\right) ^{2}-1}$. Using that $%
y_{0}^{-1}=$ $k_{B}T/\Delta <<1$, we obtain,
\begin{equation*}
I_{AF}(0,T)\approx \frac{2\sqrt{2}\Delta ^{2}}{D^{3/2}}e^{-\frac{\Delta }{%
k_{B}T}}\left[ \left( \frac{k_{B}T}{\Delta }\right) ^{\frac{1}{2}}+\frac{5}{4%
}\left( \frac{k_{B}T}{\Delta }\right) ^{\frac{3}{2}}\right]
\end{equation*}%
the first term in this expansion yields the result of Ref.\cite{yamada}.
Furthermore, it is easy to check that in the gapless case the temperature
dependence of the resistivity due to linearly dispersive magnons is, $\rho
(T)\propto T^{2}$ \cite{yamada}. Eq. \ref{antiferrogap} allows to obtain
from the measured $\rho (T)$, the absolute value of the spin-wave gap. If $%
\rho (T)$ is measured for different pressures it yields the trend of the
stiffness $D$ and gap $\Delta $ as a function of this parameter, for
example, when approaching a QCP. In a recent analysis of a magnetic heavy
fermion driven to a QCP by pressure, we obtained that $\Delta \propto
(P_C-P) $, vanishing with pressure at the same critical pressure $P_C$ that $%
T_N(P)$ vanishes while the stiffness $D$ remains unrenormalized
\cite{suzana}. It would be interesting to confirm these results by
direct neutron scattering experiments \cite{23,24}.

\section{Zero temperature field induced transition}

We now turn our attention to the next topic of this Report which
deals with the phase transition induced in $AF$ heavy fermions by an
\emph{external magnetic field}. In this case it is useful to start
in the paramagnetic
phase, with $H>H_{C2}(T=0)$ (see Fig. \ref%
{fig1}). The results below apply in general even if the external
field is applied perpendicularly to the easy axis. The spin-wave
spectra for $H > H_{C2}$ is ferromagnetic like, with a dispersion
relation given by \cite{keffer,callen},
\begin{equation}
\omega (\mathbf{q})=\mu H-S[J(0)-J(\mathbf{q})]
\end{equation}%
where the minus sign takes into account explicitly the antiferromagnetic
nature of the exchange interactions ($J$). In a simple cubic lattice, the
minimum of the spin-wave spectrum occurs for $\mathbf{q}=\mathbf{Q}=(\pi
/a,\pi /a,\pi /a)$. For simplicity we have considered only the isotropic
part of the exchange coupling \cite{keffer,yamada,callen}. Since $\omega
(Q)=\mu H-2SZJ$, the condition $\omega (Q)\equiv 0$ defines the critical
field $H_{C2}(T=0)=2SZJ/\mu $, below which the spin-wave energy becomes
negative signalling the entrance of the system in the $AF$ or spin-flop
phase. Above, $Z$ is the number of nearest neighbors. Finally, writing $%
\mathbf{q}=\mathbf{Q}+\mathbf{k}$ and expanding for small $k$, the spin-wave
dispersion relation can be obtained as,
\begin{equation}
\omega (k)=\mu (H-H_{C2})+Dk^{2}
\end{equation}%
where the spin-wave stiffness at $T=0$, $D=SJa^{2}$ ($a$ is the
lattice parameter). The temperature dependence of $H_{C2}(T)$ is
determined
by that of the spin-wave stiffness $D(T)$ \cite%
{keffer,callen}. This generally results in a critical line varying
with temperature as, $H_{C2}(T)=H_{C2}(T=0)-bT^{3/2}$.

We can obtain the magnon contribution to the electrical
resistivity in the presence of the external magnetic field as,
$\rho (T)\propto I_{F}(H,T)$, where in three dimensions (3d)
\cite{yamada},
\begin{equation}
I_{F}(H,T)=\left( \frac{k_{B}T}{\mu H_{C2}}\right)
^{3/2}\int_{y_{0}}^{\infty }dy\frac{y\sqrt{y-y_{0}}}{\sinh ^{2}\frac{y}{2}}
\label{ferro}
\end{equation}%
with $y_{0}=\Delta /k_{B}T$. The gap $\Delta =\mu (H-H_{C2})$ is the control
parameter for the zero temperature field induced phase transition.

\begin{figure}[th]
\centering \includegraphics[angle=0,scale=0.90]{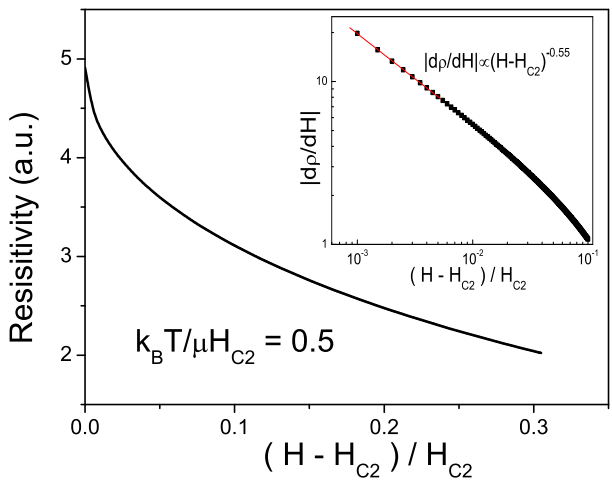}
\caption{(Color online) Field dependence of the electron-magnon resistivity
for a fixed temperature in region II of the phase diagram of Fig.\protect\ref%
{fig1}. The inset shows the field dependence of the modulus of the
derivative $d\protect\rho /dH$, which diverges as a power law when $H$
approaches $H_{C2}$.}
\label{fig2}
\end{figure}

We can distinguish two regions, separated by the crossover line
$T^{\times }=(\mu /k_{B})(H-H_{C2})$, in the phase diagram of
Fig.\ref{fig1}. This line is fully determined by the value of the
magnetic moment $\mu $ and by $H_{C2}$. In region II, for
$T>>T^{\times }$ and in particular for $H=H_{C2}$, the dominant
contribution to the resistivity is due to gapless magnons and
given by, $\rho \propto
(k_{B}T/\mu H_{C2})^{3/2}$. In this region, for a fixed temperature $%
T>>T^{\times }$, the resistivity as a function of magnetic field obtained
from Eq.\ref{ferro} is shown in Fig.\ref{fig2} . The resistivity has a cusp
at $H_{C2}$. In fact $d\rho /dH\propto -(H-H_{C2})^{-1/2}$ and diverges to $%
-\infty $ as $H\rightarrow H_{C2}$ \cite{yamada}.

In region I, for $k_{B}T<<\mu (H-H_{C2})$ the resistivity is calculated as
in the previous section and is given by,
\begin{equation}
\rho \propto \left( \frac{k_{B}T}{\mu H_{C2}}\right) ^{3/2}\left( 1+\frac{2}{%
3}\frac{\mu (H-H_{C2})}{k_{B}T}\right) e^{-\frac{\mu (H-H_{C2})}{k_{B}T}}
\label{ferrogap}
\end{equation}

The thermodynamic properties of the system close to $H_{C2}$ can
be easily obtained in the soft mode scenario. They are essentially
those of a system of non-interacting bosons. The temperature and
field dependent contribution of the spin-wave excitations to the
specific heat $C_{V}$ can be shown to have a form similar to that
of the electrical
resistivity. We get, $C_{V}\propto $ $I_{F}(H,T)$, with $%
I_{F}(H,T)$ given in Eq.\ref{ferro}. Consequently, in region II, which
includes the line $H=H_{C2\text{,}}$, the specific heat has a simple power
law dependence, $C_{V}\propto (k_{B}T/\mu H_{C2})^{3/2}$. In region I, the
field and temperature dependence of $C_{V}$ is the same as that of the
resistivity given in Eq. \ref{ferrogap}. The magnetization at the critical
field $H_{C2}$ decreases with a $T^{3/2}$ power law. In region I for $%
H>H_{C2}$ it decreases exponentially due to the gap, $\Delta =\mu
(H-H_{C2})$.

\section{Two dimensional case}

\begin{figure}[th]
\centering \includegraphics[angle=0,scale=0.75]{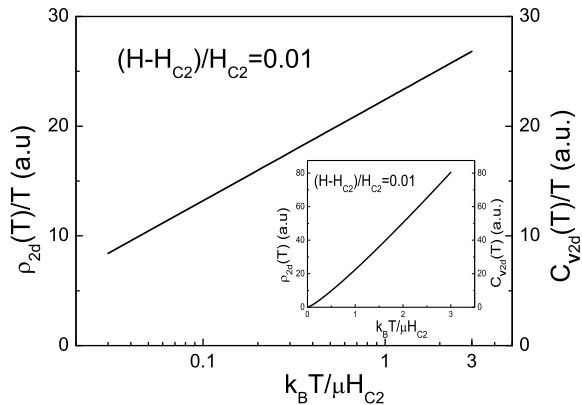}
\caption{Semi-logarithmic plot of the resistivity $\protect%
\rho$ and specific heat $C_V$ divided by T in 2d. Inset shows $\protect\rho$
and $C_V$ versus T.}
\label{fig3}
\end{figure}

Many heavy fermions systems due to their anisotropy in the
crystalline structure have a regime close to their QCP where the
critical fluctuations have a two-dimensional character
\cite{review}. For this reason it is interesting to generalize the
results obtained above for the case of two dimensions (2d). In $2d$
the relevant integral for the resistivity can be evaluated and we
get,
\begin{equation}
\rho _{2d}(h,t)\propto t\,\left( \frac{2h}{t}\coth \frac{h}{2t}-4\ln 2-4\ln
\sinh \frac{h}{2\,t}\right)
\end{equation}%
where $h=(H-H_{C2})/H_{C2}$ and $t=k_{B}T/\mu H_{C2}$. In the limit $%
(h/t)\rightarrow 0$, this reduces to
\begin{equation}
\rho _{2d}(h,t)=\rho _{2d}^{0}(1+\ln \frac{t}{h})t  \label{r2d}
\end{equation}%
This resistivity varies nearly linear in temperature (Fig.\ref{fig3}) but for $%
H\rightarrow H_{C2}$ ($h\rightarrow 0$) it is not defined for
finite temperatures. This is related to the fact that for an
isotropic system there is no long range order in two dimensions at
any finite temperature due to the creation of an infinite number
of magnons. The systems we are considering are in fact three
dimensional, although they may exhibit in some region of their
phase diagram a $2d$ behavior. However, this is just a regime
which will cross over to $3d$ sufficiently close to the
transition. Consequently this divergence of $\rho _{2d}(0,t)$
should not interfere with the actual observation of a nearly
linear temperature dependence of the resistivity of real systems
in some range of region II. In $2d$ the spin-wave contribution to
the specific heat, as before, has a similar field and temperature
dependence of the resistivity due to
electron-magnon scattering. We find in this case ($%
(h/t)\rightarrow 0$) that in region II (Fig.\ref{fig3}),
\begin{equation*}
\frac{C_{V}(h,t)}{T}\propto a(1+\ln \frac{t}{h}).
\end{equation*}

The decrease of the magnetization with temperature in a fixed field
in this region also has a field and temperature dependence similar
to Eq. \ref{r2d}.

Finally, in region I the specific heat, magnetization and
resistivity have an exponential term, as in Eq. \ref{ferrogap}.
Only the temperature dependence of the pre-factor is dependent on
dimensionality. This makes it difficult to distinguish a $2d$ from
a $3d$ regime in this region, either in thermodynamic or transport
measurements.

\section{Conclusions}

We have obtained the electrical and thermodynamic properties of a
metallic antiferromagnetic system due to spin-wave modes close to
a zero temperature field induced transition. The existence of
these excitations close to this phase transition is a general
phenomenon, depending only on the appearance of a symmetry-broken
state below $H_{C2}$. The resistivity is determined by the
scattering of quasi-particles by ferromagnetic-like magnons which
become gapless at the transition. The  contribution of the
spin-wave modes to the thermodynamic and transport properties is
distinct in different regions of the phase diagram. In region I,
this is exponentially suppressed by the gap $\Delta =\mu
(H-H_{C2})$, but in region II these modes are effectively gapless
and give rise to important contributions. Notice that these
contributions should be considered \emph{in addition to those of
other quasi-particles.} For example, in region I, it has been
observed \cite{fieldinduced} a significant $T^{2}$ term in the
resistivity probably due to electron-electron interaction. This
temperature dependence of the resistivity is different from that
arising from spin-wave scattering which in turn is exponentially
suppressed by the gap. In region II however, the spin-wave term is
large and its power law temperature dependence is similar to that
associated with non-Fermi liquid behavior close to a QCP
\cite{11}. This will make it difficult to separate this
contribution from that of other quasi-particles. The interaction
of the latter with the spin-wave modes may even give rise to
substantial mass renormalization. The same remarks apply for the
thermodynamic properties in both regions.

The temperature dependence of the thermodynamic properties at the
critical field $H_{C2}(T=0)$ is similar to that predicted by the
scaling theory of a QCP with a dynamic exponent $z=2$ \cite{11}.
In the present case, this value of $z$ reflects the
ferromagnetic-like dispersion of the magnons and not the
antiferromagnetic nature of the transition. In the extreme
metamagnetic case of an Ising antiferromagnet in an external
longitudinal field, a renormalization group approach yields a
phase diagram where the $T=0
$ transition at $H_{C2}$ is not associated with a fixed point. The point ($%
H=H_{C2},T=0$) is just an ordinary point that flows to
($H=0,T=T_N$) as any point in the line of second order transitions
$H=H_{C2}(T)$, since the magnetic field turns out to be an
\emph{irrelevant} parameter \cite{raimundo}. Notice that the
spin-wave approach presented above is independent whether the
$T=0$ transition at $H_{C2}$ is associated or not with a fixed
point.

\acknowledgements{We thank S. Bud`ko, J. Larrea and Magda Fontes
for helpful discussions. Support from the Brazilian agencies CNPq
and FAPERJ is gratefully acknowledged. Work partially supported by
PRONEX/MCT and FAPERJ/Cientista do Nosso Estado programs.}

\end{document}